# Zeolitic Imidazolate Framework-8 offers an anti-inflammatory and antifungal method in the treatment of *Aspergillus fungus* keratitis in vitro and in vivo

Xueyun Fu[1*], Xue Tian[1*], Jing Lin[1], Qian Wang[1], Lingwen Gu[1], Ziyi Wang[1], Menghui Chi[1], Bing Yu[1], Zhuhui Feng[1], Wenyao Liu[1], Lina Zhang[1], Cui Li[1], Guiqiu Zhao[1]

[1] Department of Ophthalmology, The Affiliated Hospital of Qingdao University, Qingdao, People's Republic of China

*These authors contributed equally to this work

Correspondence: Cui Li; Guiqiu Zhao, Department of Ophthalmology, Affiliated Hospital of Qingdao University, 16 Jiangsu Road, Qingdao 266003, Shandong, People's Republic of China, yankelicui@126.com; zhaoguiqiu_good@126.com

**Abstract:**

**Background:** Fungal keratitis is a serious blinding eye disease. Traditional drugs used to treat fungal keratitis commonly have the disadvantages of low bioavailability, poor dispersion, and limited permeability.

**Purpose:** To develop a new method for the treatment of fungal keratitis with improved bioavailability, dispersion, and permeability.

**Methods:** Zeolitic Imidazolate Framework-8 (ZIF-8) was formed by zinc ions and 2-methylimidazole linked by coordination bonds and characterized by Scanning electron microscopy (SEM), X-ray diffraction (XRD), and Zeta potential. The safety of ZIF-8 on HCECs and RAW 264.7 cells was detected by Cell Counting Kit-8 (CCK-8). Safety evaluation of ZIF-8 on mice corneal epithelium was conducted using the Driaze corneal toxicity test. The effects of ZIF-8 on fungal growth, biofilm formation, and hyphae structure were detected by Minimal inhibit concentration (MIC), crystal violet staining, Propidium Iodide (PI) testing, and calcofluor white staining. The anti-inflammatory effects of ZIF-8 on RAW 246.7 cells were evaluated by Quantitative Real-Time PCR Experiments (qPCR) and Enzyme-linked immunosorbent assay (ELISA). Clinical score, Colony-Forming Units (CFU), Hematoxylin-eosin (HE) staining, and immunofluorescence were conducted to verify the therapeutic effect of ZIF-8 on C57BL/6 female mice with fungal keratitis.

**Results:** In vitro, ZIF-8 showed outstanding antifungal effects, including inhibiting the growth of *Aspergillus fumigatus* over 90% at 64 μg/mL, restraining the formation of biofilm, and destroying cell membranes. In vivo, treatment with ZIF-8 reduced corneal fungal load and mitigated neutrophil infiltration in fungal keratitis, which effectively reduced the severity of keratitis in mice and alleviated the infiltration of inflammatory factors in the mouse cornea. In addition, ZIF-8 reduces the inflammatory response by downregulating the expression of pro-inflammatory cytokines TNF-α, IL-6, and IL-1β after *Aspergillus fumigatus* infection in vivo and in vitro.

**Conclusion:** ZIF-8 has a significant anti-inflammatory and antifungal effect, which provides a new solution for the treatment of fungal keratitis.

**Keywords:**

## Introduction

Fungal keratitis is a severe corneal infection disease. The most common causes of fungal keratitis are *Fusarium* and *Aspergillus*.[1,2] Once filamentous fungi conidia settle on the damaged corneal epithelium, they generate hyphae, extracellular enzymes, and toxins to disrupt the corneal endothelium.[3] Recent research reported that the annual increase in patients with fungal keratitis may exceed one million, with the majority occurring in developing countries.[4-6] Due to fungal keratitis, they may experience visual impairment or even blindness.[7-9] Therefore, there is a growing interest in how to treat fungal keratitis in an affordable and efficient manner.

Drug therapy is the primary method for treating fungal keratitis.[10] However, a major issue with commonly used drugs like natamycin is their low bioavailability and poor dispersibility.[11-14] While the extensively researched composite nanomaterials have solved the above problem,[15-17] most of their nanocarriers only have delivery effects, and the vector itself often has no biological function. The investigation by Jian Yu et al. utilized UiO-66 nanoparticles as carriers for moxifloxacin, a bacteria-targeting peptide ubiquitin, and ROS-responsive poly (ethylene glycol)-thioketal. This combination was employed to treat endophthalmitis effectively.[15] Nanocarriers with their own biological functions may allow this drug delivery system to play a better therapeutic effect. In addition, it is characterized by high water solubility, good dispersibility, and biocompatibility. Therefore, it would be beneficial to search for a nanocarrier that can both deliver and exert antimicrobial and anti-inflammatory effects.

Zeolitic Imidazolate Framework-8 (ZIF-8), a class of nanomaterial formed by metal-organic coordination bonds, has a high surface area and exhibits excellent chemical-thermal stability.[18-21] Furthermore, it exhibits notable attributes such as high solubility in water, excellent dispersibility, and biocompatibility.[22] ZIF-8 has garnered evident attention in the drug delivery for antibacterial and anticancer agents. Lichuan Tan et al. studied that Physcion coated with ZIF-8 could inhibit the growth of Escherichia coli and Staphylococcus aureus, and Jinguo Qiu et al. found that ZIF-8 loaded with anticancer peptides could be used to treat breast cancer.[23,24] Over the past decade, ZIF-8 has been shown to be an antibacterial agent.[25-30] However, its application in antifungal therapy remains limited. Currently, only studies have shown that ZIF-8 is a potential treatment to aquatic fungi associated with leaf litter decomposition. The findings indicated that ZIF-8 nanoparticles diminished dehydrogenase activity and markedly suppressed fungal biomass accumulation.[31] The imidazole group of ZIF-8 suggests inherent antifungal properties, but it has not been used in fungal infectious diseases. By inhibiting lanosterol 14α demethylase, imidazoles prevent the synthesis of ergosterol, a major component of fungal cell membranes, causing cell death due to disturbed cell permeability.[32,33] Therefore, our study aims to investigate the antifungal effects of ZIF-8 by examining the morphology and growth of the fungi. Additionally, we sought to assess the anti-inflammatory properties of ZIF-8 by analyzing changes in proinflammatory factors in the messenger RNA and protein levels. Finally, the therapeutic effects of ZIF-8 were evaluated based on clinical scores and corneal fungal load in mice infected with fungal keratitis. The objective is to identify a more efficacious therapeutic agent for treating fungal keratitis and to explore novel avenues for the subsequent utilization of ZIF-8 as either a carrier or a drug in fungal infectious diseases.

In this paper, ZIF-8 was synthesized, and the antifungal and anti-inflammatory effects of ZIF-8 were verified in vivo and in vitro. In vitro assessment, we confirmed its toxicity and examine the effect of ZIF-8 to TNF-α, IL-1β and IL-6. We also verified its inhibitory ability against *Aspergillus fumigatus* through its effect on hyphae growth inhibitory, hyphae destruction, biofilm growth destruction and cell wall growth inhibitory, and cell membrane destruction. In vivo evaluations, after treatment with ZIF-8, the keratitis was alleviated, the fungal load was reduced, the tissue edema was alleviated, and the inflammatory cell infiltration was reduced, which effectively reduced the infiltration of inflammatory factors in the mouse cornea (Scheme 1). In short, this paper researched the anti-fungal and anti-inflammary properties of ZIF-8 and provides a new field and idea for the treatment of fungal keratitis.

# Material and methods

## *Material*

Zn (NO$_3$)$_2$·6H$_2$O was bought from Aladdin (Shanghai, China). Methanol and 2-methylimidazole were purchased from Macklin (Shanghai, China). The phosphate-buffered saline (PBS) solution was purchased from BI Corporation (Beit Haemek, Israel). NATA was ordered by MedChemExpress (MCE; NJ, USA). Cell Counting Kit-8 (CCK-8) was acquired from MCE. Elisa mouse TNF-α kit, Elisa mouse IL-6, and Elisa mouse IL-1β kit were bought from Biolegend (USA). PI (DA0022) was procured from Leagene Biotechnology (China). Calcofluor white (18909) was obtained from Sigma-Aldrich (USA). Crystal violet and RIPA were obtained from Solarbio (China).

## *Synthesis and characterization of ZIF-8*

The synthesis process was commenced by preparing a methanol solution (1#) of Zn (NO$_3$)$_2$·6H$_2$O with 5.58 grams in 150 milliliters. Simultaneously, 6.16 grams of 2-methylimidazole was added into a separate beaker containing 150 milliliters of methanol, ensuring thorough homogenization (2#). Subsequently, the 1# and 2# were gradually poured in under continuous stirring. The composite was agitated comprehensively for a duration of 24 hours. The ZIF-8 product is obtained by executing centrifugation, followed by ethanol washing, and concluding with vacuum drying at 60°C for 10 hours. The stoichiometric ratio of 2-methylimidazole to zinc was 4:1 throughout the synthetic procedure. After the synthesis process, the synthesized ZIF-8 powder was uniformly dispersed onto a conductive adhesive tape (C-tape), followed by the removal of any unattached powder using an earwash ball and subsequently examined under a Scanning electron microscope (SEM; JSM-7001F; Acceleration voltage: 500V-30 kV, resolution: 1.0 nm; JEOL, Tokyo, Japan), and EDS-mapping was used to analyze element composition of these powder. X-ray diffraction (XRD; Smart Lab ⩾ 3 KW; ⩾ 60 kV; ⩾ 60 mA, Rigaku, Japan) was used for qualitative analysis of synthesized powders. Zeta potential (DLS-Zeta potential, particle size range: 3.8 nm-100 um,

Nano ZSE, Malvern, Britain) was to check the stability of the material, we tested the material by dispersing it uniformly in deionized water.

## *CCK-8*

To assess the impact of ZIF-8 on cell viability, the CCK-8 assay was employed to measure the viability of human corneal epithelial cells (HCECs, gained from Xiamen Eye Center of Xia Men University) and mice mononuclear macrophage cells (RAW 264.7 cells, gained from Academy of Sciences in Shanghai) after ZIF-8 treatment. The study on HCECs was approved by the Ethics Review Committee of the Affiliated Hospital of Qingdao University, Qingdao, Shandong, China (QYFYWZLL 28813). HCECs and RAW 264.7 cell suspensions ($1\times10^4$ cells/mL) were separately seeded in a 96-well plate at 100μL per well and cultured for approximately 24 hours at 37°C with 5% $CO_2$. Then ZIF-8 was diluted using a serial dilution method to achieve different concentrations (2, 4, 8, 16, 32, 64, 128, 256 μg/mL), which were added to the culture medium with each concentration group replicated 6 times. The control group was an equal volume medium with HCECs or RAW 264.7 cells. After the 24-hour incubation period, 10 μL of CCK-8 reagent was added to each well, followed by a 2-hour incubation. The Optical density (OD) values were measured using enzyme labeling at 450 nm. These experimental procedures were conducted to thoroughly elucidate the potential impact of ZIF-8 on cellular viability.

## *Draize Test*

Moreover, 5 μL of ZIF-8 at 640 μg/mL was instilled into the conjunctival sac of the right eye of mice (female 8-week-old C57BL/6 mice were supplied by Pengyue Co. Ltd.), while the left eye was treated with PBS solution as the control group. On days 1, 3, and 5 post ZIF-8 administration, the corneal surface of the mice was observed under cobalt-blue light using a slit lamp to detect fluorescein sodium staining. The scoring criteria for fluorescein staining (CFS) were determined according to the guidelines outlined by the OECD,[34,35] providing a quantitative evaluation of the severity of adverse reactions.

Information Classification: General

## *Minimum Inhibitory Concentration (MIC)*

We utilized Sabouraud liquid medium containing spores of *Aspergillus* (strain 3.0772, China General Microbiological Culture Collection Center), at a concentration of $3×10^5$ CFU/mL to dilute ZIF-8. ZIF-8 was diluted into different concentrations (2, 4, 8, 16, 32, 64, 128, 256 μg/mL). The control group was Sabouraud liquid medium with $3×10^5$ CFU/mL spores. Subsequently, the samples of various concentrations were added to a 96-well plate with each sample replicated 6 times, followed by a 36-hour incubation. Finally, the OD values at 540 nm were measured using an enzyme labeling.

## *Biofilm inhibition experiment*

The biofilm experiment is based on the steps of MIC experiment. The surface of the hyphae is removed, each hole is gently rinsed with PBS for 3 times, and then dried at room temperature. Then anhydrous ethanol is added to fix the biofilm, and then rinsed with deionized water. After air drying at room temperature, crystal violet dye was added for 15 minutes, crystal violet was removed, and the unbound crystal violet was fully rinsed with sterilized PBS, then decolorized with 95% ethanol for 10 minutes, and 100 μL of the mixture was absorbed and transferred to a new 96-well plate. The OD values at 570 nm were measured by enzyme labeling.

## *Propidium Iodide (PI) testing*

To explore the effect of ZIF-8 on fungal viability, live/dead staining was conducted. 1 mL of *Aspergillus* spore suspension (concentration of $10^5$ CFU/mL) was added to each well of a 6-well plate and cultured for 24 hours in a constant temperature incubator (37°C, 5% $CO_2$). Upon forming a thin layer of hyphae on the surface, the hyphae were collected, washed 3 times, and centrifuged at 12000 rpm for 10 minutes. Then, the hyphae were treated with ZIF-8 (32 μg/mL and 64 μg/mL) Sabouraud liquid medium and natamycin (NATA) (8 μg/mL) Sabouraud liquid medium. The mycelia of the control group were added with Sabouraud solution. After another 24 hours of incubation, propidium iodide (PI) dye (50 μg/mL, 1 mL per well) was introduced. Following a 15-minute duration at room temperature without light, fluorescent images were captured under green excitation light using a fluorescence microscope. The entire procedure was repeated 3 times to ensure reliability.

## *Calcofluor White Staining*

A calcofluor white staining experiment was conducted to investigate the impact of ZIF-8 on fungal cell wall integrity. ZIF-8 (16, 32 and 64 μg/mL) and NATA (8 μg/mL) were introduced into the Sabouraud liquid medium containing spores ($1\times10^5$ CFU/mL). Sabouraud liquid medium containing spores ($1\times10^5$ CFU/mL) was added to control group. The resulting mixture was added to a 12-well plate (1 mL per well) and placed in a constant temperature incubator (37°C, 5% $CO_2$) for 24 hours. The mixed solution was added to a 12-well plate (1mL per well) and incubated for 24 hours at 37°C with 5% $CO_2$. Subsequently, fungal hyphae were collected, washed 3 times with PBS, and stained with calcofluor white dye. After a 15-minute incubation in dark, observations of good conditions were made using a fluorescence microscope, and corresponding images were captured.

## *Hyphae scanning electron microscopy*

To assess the structural alterations in fungal hyphae after ZIF-8 treatment, Scanning Electron Microscopy (SEM) was used to observe the drug-treated hyphae. A spore suspension ($3\times10^5$ CFU/mL) of *Aspergillus* was added at 1 mL per well in a 12-well plate and incubated for 24 hours in a constant temperature chamber (37°C, 5% $CO_2$). After centrifugation (12000 rpm for 10 minutes), the Sabouraud liquid medium was aspirated. A new medium containing PBS (control group) and 64μg/mL of ZIF-8 was introduced. Subsequent incubation occurred for an additional 24 hours in the constant temperature chamber. The fungal hyphae were then collected, fixed with 2.5% glutaraldehyde, and left in a refrigerator at 4°C for 24 hours. After critical point drying and gold sputter coating, the morphological features of the fungal hyphae were observed under electron microscopy.

## *Establishment of animal disease mode*

The mice were anesthetized with 8% chloral hydrate (0.4 mL/kg). Using a 1 mL empty needle gently lifted the near central part of the mice cornea, which emerged as a punctate pattern with a depth reaching the stromal layer and further extended laterally to form a tunnel. A 2 μL suspension of fungal spores ($3 \times 10^7$ CFU/mL) was injected along the tunnel into the stromal layer of the right eye while the left eye remained untreated.[36,37]

The method of mice euthanasia is as follows: the mice were placed in a carbon dioxide anesthesia tank, waited for loss of consciousness, and then killed by cervical dislocation.

## *Corneal colony count*

Corneas from mice infected with *Aspergillus fumigatus* for 3 days were collected in pairs. The experimental group was ZIF-8 treatment group, and the control group was PBS treatment group. They were homogenized in 200 μL of sterile PBS using ultrasound and then cultured on Sabouraud agar medium. After 36 hours of incubation (37°C), colony counts of *Aspergillus fumigatus* were recorded, and photographs were taken.

## *Hematoxylin and eosin (HE) staining*

Mouse eyeballs were excised and fixed with 4% paraformaldehyde at 4°C for 3 days. The mouse eyeballs were then cut into 10μm thick histological sections after paraffin embedding. After deparaffinization in xylene and dehydration with an ethanol concentration gradient, the tissue sections were stained with hematoxylin and eosin. They were observed and photographed under a microscope with a magnification of 400x.

## *Immunofluorescence staining (IF)*

The mice eyeballs were taken and placed in an optimal cutting temperature compound (OCT) embedding agent, and the slices (10 μm) of the eyeballs were cut by freezing microtome after quick freezing with liquid nitrogen. After baking (37°C, 8 hours), the specimen was fixed with methanol (4°C, 30 minutes) and soaked with PBS (room temperature, 5 minutes, 3 times). After drying, the serum from the second antibody was added and sealed (room temperature, 30 minutes), and the sealing liquid was absorbed and dried. The primary antibody was incubated (4°C, overnight) and soaked with PBS (room temperature, 5 minutes, 3 times). The second antibody was incubated (away from light, 37°C, 1 hour) and soaked with PBS (away from light, room temperature, 5 minutes, 3 times). DAPI incubation (away from light, room temperature, 10 minutes), PBS immersion (away from light, room temperature, 5 minutes, 3 times). After drying away from light, the film is sealed with an anti-fluorescence attenuating tablet. The fluorescence of neutrophils was viewed and photographed under a fluorescence microscope (400x).

## *Quantitative Real-Time PCR Experiments*

RAW 264.7 cells were inoculated in 12-well plates and allowed to grow to 80%-90% confluency. Subsequently, they were stimulated with inactivated *Aspergillus fumigatus* for 1 hour, and then treated with ZIF-8 (32 μg/mL) for 7 hours after stimulation. Total ribonucleic acid (RNA) was collected for quantitative real-time polymerase chain reaction (qRT-PCR). The qRT-PCR procedure examined the nucleic acid levels of inflammatory factors IL-6, TNF-α, IL-1β, TLR-4, HO-1, and NLRP3. The mice primers sequence of these inflammatory factors can be found in Table 1.

## *Enzyme-linked immunosorbent assay (Elisa)*

The cornea was harvested from mice and placed in PBS containing phenylmethylsulfonyl fluoride (PMSF). The corneal tissues were then pulverized using a Tissue Lyser. After lysing and centrifugation of the corneal tissues, the supernatant was collected for Elisa kit analysis. For RAW 264.7 cells, the supernatant were collected directly for testing. Protein concentration was determined by measuring absorbance at 450 nm and 570 nm.

## *Statistical analysis*

The difference between the two groups was statistically significant using the student's t-test. The significance among the three or more groups was tested by one-way analysis of variance. All data are expressed as mean ± standard deviation (SD) in statistical analysis. These experiments were repeated more than 3 times.

# Results and discussion

## *Synthesis and characterization of ZIF-8*

ZIF-8 was synthesized via the room temperature synthesis.[38] In this way, 2-methylimidazole is linked to zinc ions by a coordination bond. It is worth noting that this method is simple and fast. SEM (Figure 1A-B) and energy dispersive spectrometer (EDS) mapping (Figure 1C-F) showed that ZIF-8 had a polyhedral structure, and its constituent elements included Zn, N, and C. The size of 92 ZIF-8 particles were measured, exhibiting a size range from 128 nm to 411 nm, with varying particle dimensions. Moreover, the average particle diameter of ZIF-8 was measured to be about 279 nm (Figure 1G). The value of Zeta potential showed the average potential of ZIF-8 particles is about +32 mV (Figure 1H), indicating that the solution system is relatively stable. Powder X-ray diffraction (PXRD) diagrams (Figure 1I) showed the characteristic peak (011), (002), (112), (022), (013), (222), and the outcome is consistent with the XRD results of ZIF-8 in previous studies.[21,39,40] In summary, we synthesized a stable ZIF-8.

## *Toxicity of ZIF-8 in vivo and in vitro*

As the outermost layer of the cornea, HCECs are highly resistant to microbial invasion and act as a protective barrier in the deeper layers of the cornea.[41,42] Macrophages play a role in killing pathogens and protecting the cornea in the early stage of FK infection.[43,44] Cytotoxicity assays using the CCK-8 method were conducted in vitro to evaluate the impact of varying ZIF-8 concentrations (2, 4, 8, 16, 32, 64, 128, 256 μg/mL) on RAW 264.7 cells and HCECs. Moreover, the ocular toxicity of ZIF-8 at a strength of 640 μg/mL was examined in vivo using the Draize corneal toxicity assay in C57BL/6 mice. The results of the CCK-8 cell viability assay indicated that ZIF-8 exhibits safety on RAW 264.7 cells and HCECs at low concentrations (Figure 2). No toxicity was observed for RAW 264.7 cells at 2, 4, 8, 16, 32 μg/mL. Noticeable toxicity was noted for RAW 264.7 cells at 64, 128, 256 μg/mL (Figure 2A, $p < 0.0001$). While no toxicity was observed for HCECs at 2, 4, 8, 16, 32, 64 μg/mL, cytotoxicity was noted for HCECs at 128, 256 μg/mL (Figure 2B, $p < 0.0001$). Thus, the dosage levels of ZIF-8 for subsequent experiments involving RAW 264.7 cells and HCECs were selected 32 μg/mL and 64 μg/mL, respectively.

The Draize corneal toxicity test was employed to assess whether ZIF-8 and 5% NATA caused damage to mice corneas after ocular application. The damage of corneal epithelium was recorded under cobalt blue light using a slit lamp. Compared to the PBS group, there was no epithelial damage on days 1, 3, and 5 following ocular administration of ZIF-8 and 5% NATA (Figure 2C). Draize scores showed that 640 μg/mL ZIF-8 and 5% NATA had no damage to the corneal epithelium of mice on days 1, 3, and 5 (Figure 2D). Consequently, we selected 640 μg/mL ZIF-8 and 5% NATA for vivo experiments.

ZIF-8 is composed of imidazole and zinc ions.[21] Zinc ions are trace elements within the human system, while imidazole represents an amino acid constituent that can be metabolized by the body.[45] Previous experiments have shown that ZIF-8 has been widely used as a carrier in the biomedical field,[20,46,47] such as ZIF-8 exhibited cell viability exceeding 90% in human cervical cells at a concentration of 30 μg/mL.[40] It is highly biocompatible with therapeutic drugs/enzymes in combination with the cells and has been used in animal experiments, which is in line with our experimental results.

## *Anti-inflammatory effect of ZIF-8 in vitro*

As typical inflammatory markers, elevated levels of IL-6, IL-1β, and TNF-α can represent an enhanced inflammatory response.[48-50]

Macrophages (RAW 264.7 cells) were stimulated with inactivated hyphae to mimic fungal exposure, followed by a 1 hour incubation period. After stimulation, medication was administered and maintained for 7 hours. Subsequently, quantitative polymerase chain reaction (qPCR) was employed to assess changes of inflammatory factors and related factors. Our results showed that there was a considerable rise in inflammatory markers IL-6, IL-1β and TNF-α in RAW 264.7cells stimulated by *Aspergillus fumigatus* hyphae. Conversely, the expression of IL-6 (Figure 3A, $p < 0.0001$), IL-1β (Figure 3B, $p < 0.0001$), and TNF-α (Figure 3C, $p < 0.0001$) decreased in the treatment group with ZIF-8 (32 μg/mL), indicating the anti-inflammatory effect of ZIF-8.

Previous studies have shown that activation of HO-1 can inhibit the expression of pro-inflammatory factors.[51,52] The qPCR results indicated that the expression of HO-1 increased in RAW 264.7 cells stimulated by *Aspergillus fumigatus* hyphae after ZIF-8 treatment (Figure 3D, $p < 0.0001$).

The increase of TLR-4 can promote the release of inflammatory factors and induce an inflammatory response.[53] Following ZIF-8 treatment, the mRNA levels of TLR-4 showed an obviously decrease (Figure 3E, $p < 0.001$) compared to the untreated group after fungal infection. Our results showed that ZIF-8 can reduce the expression of pro-inflammatory factors.

Activation of NLRP3 can induce the production and release of the pro-inflammatory cytokine IL-1β, leading to the enhancement of the pro-inflammatory response. NLRP3 played a crucial role as an inflammasome in the inflammatory response. Compared to the fungal infection group, the ZIF-8 treatment group showed a significant decrease in mRNA levels of NLRP3 (Figure 3F, $p < 0.0001$). ZIF-8 could decrease the expression of pro-inflammatory factors. Due to ZIF-8 can diminish the content of inflammatory factors, and the reduction of inflammatory factors can lower the inflammatory response, the experimental results once again proved that ZIF-8 can lessen the inflammatory response.

Previous experiments have shown that the addition of zinc to aflatoxin B1-contaminated feed reduces the levels of the inflammatory factors TNF-α and IL-1β in the serum of roosters.[54] In addition, 2-aryl-4-bis-amide imidazoles (ABAI) containing an imidazole moiety reduced the levels of IL-6, TNF-α, and IL-1β in LPS-stimulated macrophages, and ABAI also reduced the expression of TLR-4 and NLRP3.[55] However, the anti-inflammatory properties of ZIF-8, which is widely used to anti-infective, have rarely been explored. Therefore, we demonstrated that ZIF-8 also had the effect of decreasing TLR-4, NLRP3, increasing the content of the inhibitory anti-inflammatory factor HO-1, thus confirming the anti-inflammatory ability of ZIF-8.

## *Effect of ZIF-8 on Aspergillus fumigatus*

The antibacterial properties of ZIF-8 have been demonstrated, with inhibition of E. coli and S. aureus at 250 μg/mL.[27] Furthermore, Jingjing Du et al. have verified that ZIF-8 can clearly inhibit the growth of the black poplar-associated community of the aquatic fungi,[31] so we verified its effect on the growth of *Aspergillus fumigatus*.

The MIC experiment results revealed (Figure 4A) that ZIF-8 exhibited notable inhibition of *Aspergillus fumigatus* growth at 16 μg/mL ($p < 0.0001$). Moreover, the antifungal effect presented a concentration-dependent relationship, with over 90% inhibition recorded at 64 μg/mL ($p < 0.0001$). Biofilm formation is an important factor contributing to *Aspergillus fumigatus* resistance. The experiment suggested that ZIF-8 could inhibit biofilm growth at 16 μg/mL ($p < 0.0001$), and its resistance against fungal biofilms was dose-dependent (Figure 4B). The fluorescence intensity was also dependent on dosage, which proved the evident disruptive effect of ZIF-8 on fungal cell membranes. The integrity of *Aspergillus fumigatus* cell walls was evaluated through calcofluor white staining experiments (Figure 5B). The intensity of blue fluorescence in the images signified that the integrity of the cell wall decreased with the increasing concentrations of ZIF-8.

To directly evaluate the disruptive impact of ZIF-8 on hyphae, scanning electron microscopy (SEM) was conducted on *Aspergillus fumigatus* hyphae treated with ZIF-8 (Figure 4C-D). The SEM figures disclosed evident alterations in the smoothness and integrity of the hyphae in the ZIF-8-treated group compared to the normal group, suggesting distinct disruption caused by ZIF-8 (64 μg/mL) to *Aspergillus fumigatus* hyphae.

The results of the PI staining experiment (Figure 5A) showed that when the concentration of ZIF-8 reached 32 μg/mL, the PI dye bound to the nuclei of *Aspergillus fumigatus* cells, emitting red fluorescence. The fluorescence intensity also exhibited a dose-dependent relationship, which proved the evident disruptive effect of ZIF-8 on fungal cell membranes. At the same time, we also investigated the effect of natamycin, the clinical first-line drug for fungal keratitis, on the destruction of cell membranes, which, as a polyene drug, also plays a fungicidal role by destroying fungal cell membranes.[56] The experimental results showed that its cell-damaging effect was obvious at 8 μg/mL.

By inhibiting lanosterol 14α-demethylase, azoles prevent the synthesis of ergosterol, a major component of fungal cell membranes, causing cell death due to disturbed cell permeability.[57,58] In addition, positively charged $Zn^{2+}$ tends to adhere to the surface of microbial cell walls, and electrostatic interactions can occur between $Zn^{2+}$ and cell walls.[59] The above experimental results and theories proved that ZIF-8 inhibits the growth of *Aspergillus fumigatus* and the formation of fungal cell walls.

## *The therapeutic effect of ZIF-8 on mice Aspergillus fumigatus keratitis*

Due to its high specific surface area and good stability, ZIF-8 has been widely utilized as a nanocarrier for targeted therapy of bacterial diseases and photodynamic therapy.[60,61] However, its therapeutic potential against fungal diseases remains unexplored. Ding et al. developed a pH-responsive citrulline delivery system using zeolitic imidazolate framework-8 (ZIF-8) and investigated its fungicidal activity against M. oryzae, B. dothidea, and F. oxysporum.[17] Nevertheless, their study did not address the direct fungicidal effect of ZIF-8 itself. Limited research has been conducted on the application of ZIF-8 in the treatment of fungal infectious diseases. Previous experiments have demonstrated the inhibitory effect of ZIF-8 on the growth of *Aspergillus fumigatus*. Consequently, further investigation is warranted to ascertain the therapeutic efficacy of ZIF-8 against fungal keratitis. In addition, we compared its therapeutic efficacy with the clinical first-line drug natamycin (NATA) eye drops.

The successful establishment of mouse models of *Aspergillus fumigatus* keratitis was achieved through stromal injections. The infected eyes were given ZIF-8 and NATA, respectively, and the uninfected eyes were given PBS. This procedure was repeated 4 times a day for 5 consecutive days.The mice corneas were photographed on days 1, 3, and 5 post-infections (Figure 6A) with a slit-lamp camera. On the 3 day after infection, in the PBS group, the cornea was cloudy and the iris was not visible in the cloudy area, and the corneal neovascularization. The corneas of the ZIF-8 and NATA groups were cloudy, but iris can be seen through the cloud. However, the NATA group had significant neovascularization compared to the ZIF-8 group. On the 5 day post-infection, the severity of corneal clouding was lower in the ZIF-8-treated group compared to the PBS and NATA-treated groups. Additionally, the ZIF-8 group exhibited significantly fewer instances of iris neovascularization compared to the PBS and NATA-treated groups. According to the clinical scores (Figure 6B), the efficacy of ZIF-8 treatment was apparently better than that of PBS management ($p < 0.001$) and NATA treatment ($p < 0.05$). In the treatment of mice fungal keratitis,

ZIF-8 has demonstrated superior therapeutic efficacy over NATA, attributed to its enhanced dispersibility and corneal permeability.

Corneal homogenates collected on the third day post-infection were cultured on Sabouraud agar medium to assess fungal loads, the corneal homogenates from the PBS and NATA intervention group cultured a large number of *Aspergillus fumigatus* colonies (Figure 6C). While the number of *Aspergillus fumigatus* colonies in the ZIF-8 care group was obviously lower than in the PBS (Figure 6D, $p < 0.001$) and NATA (5%) treatment group ($p < 0.001$). The findings from the experiments demonstrated that ZIF-8 outperformed NATA in treating fungal keratitis.

## *Pathological changes of cornea in mice infected with Aspergillus fumigatus keratitis after ZIF-8 treatment*

Histological changes can reflect the therapeutic effect of ZIF-8 on fungal keratitis in mice. The results of HE staining attested that on the third day after infection, the infiltration of inflammatory cells in the corneal and stroma layers was remarkably reduced in the ZIF-8 therapeutic group compared to the control group and the NATA treatment group (Figure 7A). This result indicates that ZIF-8 can improve the keratitis caused by *Aspergillus fumigatus*.

Neutrophils play an important role in the Inflammation process of fungal keratitis, several studies have demonstrated high levels of neutrophil producted pro-inflammatory and chemotactic cytokines.[62,63] Our results of corneal immunofluorescence localization in mice showed that neutrophil infiltration was less in the ZIF-8 treated group than in the control group and in the NATA group (Figure 7B). This result displays that ZIF-8 can mitigate the inflammatory response of the cornea, and has therapeutic effect on fungal keratitis

## *ZIF-8 decreases proinflammatory proteins in RAW 264.7 cells and in mice corneas after fungal infection*

To further verify the anti-inflammatory effect of ZIF-8 and whether it has a pro-inflammatory effect, we used Elisa to verify the protein level of the inflammatory factors TNF-α, IL-1β, and IL-6. RAW 264.7 cells stimulated by *Aspergillus fumigatus* hyphae were treated with ZIF-8 (32 µg/mL) for 24 hours to detect changes in the expression of IL-6, IL-1β, and TNF-α protein levels. The Elisa results showed that compared with the normal group, the expression of inflammatory factors was not affected by ZIF-8, and compared with the group stimulated by addition of *Aspergillus fumigatus* hyphae, the addition of ZIF-8 could effectively reduce the expression of inflammatory factors IL-6 (Figure 8A, p<0.0001), IL-1β (Figure 8B, p<0.0001), and TNF-α (Figure 8C, p<0.0001) protein levels.

Owing to excessive inflammatory response may damage the corneal stroma and even lead to corneal perforation,[64] appropriate control of the inflammatory response caused by *Aspergillus fumigatus* infection is essential. Mice with *Aspergillus fumigatus* keratitis were treated with ZIF-8, and the changes in the expression of IL-6, IL-1β, and TNF-α proteins were also detected. The Elisa results showed that compared to the control group, ZIF-8 eye drops alone did not promote the production of inflammatory cytokines. Compared with mice with *Aspergillus fumigatus* keratitis, the application of ZIF-8 could effectively reduce the expression of the inflammatory factors IL-6 (Figure 8D, p< 0.0001), IL-1β (Figure 8E, p<0.0001), and TNF-α (Figure 8F, p<0.0001) protein expression. Taken together, this suggests that ZIF-8 can exert anti-inflammatory effects.

An unregulated inflammatory response may cause damage to the corneal stroma, resulting in heightened opacity, structural impairment, and potentially corneal perforation. Thus, the regulation of excessive inflammation is imperative to mitigate corneal injury during fungal infections. Our experimental findings demonstrate the efficacy of ZIF-8 in modulating the inflammatory factors, as evidenced by both in vivo and in vitro analyses. As a potential therapeutic agent, ZIF-8 inhibits the inflammatory response in *Aspergillus fumigatus* keratitis.

Compared with natamycin, ZIF-8 reveals good bioavailability, superb dispersion, and excellent permeability, making it more efficient.[22] Compared with other nanocarriers, the antifungal and anti-inflammatory effects of ZIF-8 make it a unique advantage in the treatment of fungal infectious diseases. [65,66] In a word, the advantages of antibacterial and anti-inflammatory properties and good bioavailability, superb dispersion, and excellent permeability make ZIF-8 play a more efficient role in the treatment of fungal keratitis.

In summary, ZIF-8 can achieve the purpose of antifungal by destroying the cell wall and membrane of *Aspergillus fungi* and inhibiting the formation of biofilm. In addition, ZIF-8 can avoid excessive inflammatory responses by reducing the levels of pro-inflammatory factors. Previous studies have shown that ZIF-8 can regulate tumor immunity by regulating the NLRP3 pathway.[67] This pathway also plays a role in fungal keratitis.[68] So, we think that ZIF-8 might have an impact on the NLRP3 pathway.

## Conclusion

This research synthesized ZIF-8 using a room-temperature stirring method. The results of the vivo and vitro experiments indicated that ZIF-8 exhibited anti-inflammatory properties while maintaining cell viability. ZIF-8 could treat the corneas of mice infected with *Aspergillus fumigatus* keratitis, reduce the inflammatory reaction, the fungal load, and the content of inflammatory cells in the cornea of fungi-infected mice. To our best knowledge, this study is the first to investigate the therapeutic effect of ZIF-8 in fungal infectious diseases and implies the new application of nanocarriers like ZIF-8 in fungal infectious diseases. In the future, we will further compare and discuss the detailed antifungal mechanism of ZIF-8 and its therapeutic effect on fungal keratitis after drug loading. This includes evaluating the effects of ZIF-8 on fungus-specific genes and fungal mitochondria, as well as investigating the therapeutic potential of ZIF-8 combined with clinical drugs such as natamycin for fungal keratitis. These studies can make a more profound study of ZIF-8, which may open up the application of ZIF-8 in more different fields.

## Ethics

Our experiments were carried out in accordance with the Basel Declaration and the ethical guidelines of the International Council for Laboratory Animal Science with the approval of the Ethics Review Commitee of the Affliated Hospital of Qingdao University, Qingdao, Shandong, China (QYFYWZLL 28813).

## Acknowledgments

This work was financially supported by the National Natural Science Foundation of China (Nos. 82171029, 81870632 and 81800800), China Postdoctoral Science Foundation (Nos. 2020M672000) and the Taishan Scholars Program (Nos. ts201511108, tsqn202103188 and tsqn201812151).

## Disclosure

The authors declare no conflict of interest.

## References

1. Sharma N, Bagga B, Singhal D, et al. Fungal keratitis: A review of clinical presentations, treatment strategies and outcomes. *Ocul Surf*. Apr 2022;24:22-30. doi:10.1016/j.jtos.2021.12.001
2. Manikandan P, Abdel-Hadi A, Randhir Babu Singh Y, et al. Fungal Keratitis: Epidemiology, Rapid Detection, and Antifungal Susceptibilities of Fusarium and Aspergillus Isolates from Corneal Scrapings. *Biomed Res Int*. 2019;2019:6395840. doi:10.1155/2019/6395840
3. Lakhundi S, Siddiqui R, Khan NA. Pathogenesis of microbial keratitis. *Microb Pathog*. Mar 2017;104:97-109. doi:10.1016/j.micpath.2016.12.013
4. Brown L, Leck AK, Gichangi M, Burton MJ, Denning DW. The global incidence and diagnosis of fungal keratitis. *Lancet Infect Dis*. Mar 2021;21(3):e49-e57. doi:10.1016/s1473-3099(20)30448-5
5. Hoffman JJ, Burton MJ, Leck A. Mycotic Keratitis-A Global Threat from the Filamentous Fungi. *J Fungi (Basel)*. Apr 3 2021;7(4)doi:10.3390/jof7040273
6. Ting DSJ, Ho CS, Deshmukh R, Said DG, Dua HS. Infectious keratitis: an update on epidemiology, causative microorganisms, risk factors, and antimicrobial resistance. *Eye (Lond)*. Apr 2021;35(4):1084-1101. doi:10.1038/s41433-020-01339-3

7. Burton MJ, Pithuwa J, Okello E, et al. Microbial keratitis in East Africa: why are the outcomes so poor? *Ophthalmic Epidemiol*. Aug 2011;18(4):158-63. doi:10.3109/09286586.2011.595041
8. Brown L, Kamwiziku G, Oladele RO, et al. The Case for Fungal Keratitis to Be Accepted as a Neglected Tropical Disease. *J Fungi (Basel)*. Oct 5 2022;8(10)doi:10.3390/jof8101047
9. Thomas PA. Current perspectives on ophthalmic mycoses. *Clin Microbiol Rev*. Oct 2003;16(4):730-97. doi:10.1128/cmr.16.4.730-797.2003
10. Sha XY, Shi Q, Liu L, Zhong JX. Update on the management of fungal keratitis. *Int Ophthalmol*. Sep 2021;41(9):3249-3256. doi:10.1007/s10792-021-01873-3
11. Mahmoudi S, Masoomi A, Ahmadikia K, et al. Fungal keratitis: An overview of clinical and laboratory aspects. *Mycoses*. Dec 2018;61(12):916-930. doi:10.1111/myc.12822
12. Liu L, Wu H, Riduan SN, Ying JY, Zhang Y. Short imidazolium chains effectively clear fungal biofilm in keratitis treatment. *Biomaterials*. Jan 2013;34(4):1018-23. doi:10.1016/j.biomaterials.2012.10.050
13. Van De Veerdonk FL, Gresnigt MS, Romani L, Netea MG, Latgé JP. Aspergillus fumigatus morphology and dynamic host interactions. *Nat Rev Microbiol*. Nov 2017;15(11):661-674. doi:10.1038/nrmicro.2017.90
14. Banshoya K, Fujita C, Hokimoto Y, et al. Amphotericin B nanohydrogel ocular formulation using alkyl glyceryl hyaluronic acid: Formulation, characterization, and in vitro evaluation. *Int J Pharm*. Dec 15 2021;610:121061. doi:10.1016/j.ijpharm.2021.121061
15. Yu J, Xu H, Wei J, Niu L, Zhu H, Jiang C. Bacteria-Targeting Nanoparticles with ROS-Responsive Antibiotic Release to Eradicate Biofilms and Drug-Resistant Bacteria in Endophthalmitis. *Int J Nanomedicine*. 2024;19:2939-2956. doi:10.2147/ijn.S433919
16. Gu L, Li C, Lin J, et al. Drug-loaded mesoporous carbon with sustained drug release capacity and enhanced antifungal activity to treat fungal keratitis. *Biomater Adv*. May 2022;136:212771. doi:10.1016/j.bioadv.2022.212771
17. Ding Y, Yuan J, Mo F, et al. A pH-Responsive Essential Oil Delivery System Based on Metal-organic Framework (ZIF-8) for Preventing Fungal Disease. *J Agric Food Chem*. Nov 29 2023;71(47):18312-18322. doi:10.1021/acs.jafc.3c04299
18. Ploetz E, Engelke H, Lächelt U, Wuttke S. The chemistry of reticular framework nanoparticles: MOF, ZIF, and COF materials. *Adv Funct Mater*. 2020;30(41):1909062.
19. Abdelhamid HN, Bermejo-Gómez A, Martín-Matute B, Zou X. A water-stable lanthanide metal-organic framework for fluorimetric detection of ferric ions and tryptophan. *Mikrochim Acta*. 2017;184(9):3363-3371. doi:10.1007/s00604-017-2306-0
20. Costa BA, Abuçafy MP, Barbosa TWL, et al. ZnO@ZIF-8 Nanoparticles as Nanocarrier of Ciprofloxacin for Antimicrobial Activity. *Pharmaceutics*. Jan 11 2023;15(1)doi:10.3390/pharmaceutics15010259
21. Park KS, Ni Z, Côté AP, et al. Exceptional chemical and thermal stability of zeolitic imidazolate frameworks. *Proc Natl Acad Sci U S A*. Jul 5 2006;103(27):10186-10191. doi:10.1073/pnas.0602439103
22. Hani Nasser A. Zeolitic Imidazolate Frameworks (ZIF-8) for Biomedical Applications: A Review. *Curr Med Chem*. Oct 27 2021;28(34):7023-7075. doi:10.2174/0929867328666210608143703
23. Tan L, Yuan G, Wang P, Feng S, Tong Y, Wang C. pH-responsive Ag-Phy@ZIF-8 nanoparticles modified by hyaluronate for efficient synergistic bacteria disinfection. *Int J Biol Macromol*. May 1 2022;206:605-613. doi:10.1016/j.ijbiomac.2022.02.097
24. Qiu J, Tomeh MA, Jin Y, Zhang B, Zhao X. Microfluidic formulation of anticancer peptide loaded ZIF-8 nanoparticles for the treatment of breast cancer. *J Colloid Interface Sci*. Jul 15 2023;642:810-819. doi:10.1016/j.jcis.2023.03.172
25. Redfern J, Geerts L, Seo JW, Verran J, Tosheva L, Wee LH. Toxicity and antimicrobial properties of ZnO@ ZIF-8 embedded silicone against planktonic and biofilm catheter-associated pathogens. *ACS Appl Nano Mater*. 2018;1(4):1657-1665.
26. Xu Y, Fang Y, Ou Y, et al. Zinc metal–organic framework@ chitin composite sponge for rapid hemostasis and antibacterial infection. *ACS Sustain Chem & Eng*. 2020;8(51):18915-18925.
27. Zhang Y, Li TT, Shiu BC, Lin JH, Lou CW. Two methods for constructing ZIF-8 nanomaterials with good bio compatibility and robust antibacterial applied to biomedical. *J Biomater Appl*. Jan 2022;36(6):1042-1054. doi:10.1177/08853282211033682

28. Kohsari I, Shariatinia Z, Pourmortazavi SM. Antibacterial electrospun chitosan-polyethylene oxide nanocomposite mats containing ZIF-8 nanoparticles. *Int J Biol Macromol*. Oct 2016;91:778-88. doi:10.1016/j.ijbiomac.2016.06.039
29. Zheng X, Zhang Y, Zou L, et al. Robust ZIF-8/alginate fibers for the durable and highly effective antibacterial textiles. *Colloids Surf B Biointerfaces*. Sep 2020;193:111127. doi:10.1016/j.colsurfb.2020.111127
30. Au-Duong AN, Lee CK. Iodine-loaded metal organic framework as growth-triggered antimicrobial agent. *Mater Sci Eng C Mater Biol Appl*. Jul 1 2017;76:477-482. doi:10.1016/j.msec.2017.03.114
31. Du J, Qv M, Qv W, et al. Potential threats of zeolitic imidazolate framework-8 nanoparticles to aquatic fungi associated with leaf decomposition. *J Hazard Mater*. Jan 5 2021;401:123273. doi:10.1016/j.jhazmat.2020.123273
32. Allen D, Wilson D, Drew R, Perfect J. Azole antifungals: 35 years of invasive fungal infection management. *Expert Rev Anti Infect Ther*. Jun 2015;13(6):787-98. doi:10.1586/14787210.2015.1032939
33. Borgers M. Mechanism of action of antifungal drugs, with special reference to the imidazole derivatives. *Rev Infect Dis*. Jul-Aug 1980;2(4):520-34. doi:10.1093/clinids/2.4.520
34. Wilhelmus KR. The Draize eye test. *Surv Ophthalmol*. May-Jun 2001;45(6):493-515. doi:10.1016/s0039-6257(01)00211-9
35. Secchi A, Deligianni V. Ocular toxicology: the Draize eye test. *Curr Opin Allergy Clin Immunol*. Oct 2006;6(5):367-72. doi:10.1097/01.all.0000244798.26110.00
36. Zhong J, Huang W, Deng Q, et al. Inhibition of TREM-1 and Dectin-1 Alleviates the Severity of Fungal Keratitis by Modulating Innate Immune Responses. *PLoS One*. 2016;11(3):e0150114. doi:10.1371/journal.pone.0150114
37. Montgomery ML, Fuller KK. Experimental Models for Fungal Keratitis: An Overview of Principles and Protocols. *Cells*. Jul 16 2020;9(7)doi:10.3390/cells9071713
38. Ta DN, Nguyen HK, Trinh BX, Le QT, Ta HN, Nguyen HT. Preparation of nano‐ZIF‐8 in methanol with high yield. *The Canadian Journal of Chemical Engineering*. 2018;96(7):1518-1531.
39. Bux H, Feldhoff A, Cravillon J, Wiebcke M, Li Y-S, Caro J. Oriented zeolitic imidazolate framework-8 membrane with sharp H2/C3H8 molecular sieve separation. *Chemistry of Materials*. 2011;23(8):2262-2269.
40. Li X, Qi M, Li C, et al. Novel nanoparticles of cerium-doped zeolitic imidazolate frameworks with dual benefits of antibacterial and anti-inflammatory functions against periodontitis. *J Mater Chem B*. Nov 28 2019;7(44):6955-6971. doi:10.1039/c9tb01743g
41. Hua X, Yuan X, Tang X, Li Z, Pflugfelder SC, Li DQ. Human corneal epithelial cells produce antimicrobial peptides LL-37 and β-defensins in response to heat-killed Candida albicans. *Ophthalmic Res*. 2014;51(4):179-86. doi:10.1159/000357977
42. Bartakova A, Kunzevitzky NJ, Goldberg JL. Regenerative Cell Therapy for Corneal Endothelium. *Curr Ophthalmol Rep*. Sep 1 2014;2(3):81-90. doi:10.1007/s40135-014-0043-7
43. Hu J, Wang Y, Xie L. Potential role of macrophages in experimental keratomycosis. *Invest Ophthalmol Vis Sci*. May 2009;50(5):2087-94. doi:10.1167/iovs.07-1237
44. Hu J, Hu Y, Chen S, et al. Role of activated macrophages in experimental Fusarium solani keratitis. *Exp Eye Res*. Dec 2014;129:57-65. doi:10.1016/j.exer.2014.10.014
45. Zhuang J, Kuo CH, Chou LY, Liu DY, Weerapana E, Tsung CK. Optimized metal-organic-framework nanospheres for drug delivery: evaluation of small-molecule encapsulation. *ACS Nano*. Mar 25 2014;8(3):2812-9. doi:10.1021/nn406590q
46. Chen H, Yang J, Sun L, et al. Synergistic Chemotherapy and Photodynamic Therapy of Endophthalmitis Mediated by Zeolitic Imidazolate Framework-Based Drug Delivery Systems. *Small*. Nov 2019;15(47):e1903880. doi:10.1002/smll.201903880
47. Cheng C, Li C, Zhu X, Han W, Li J, Lv Y. Doxorubicin-loaded Fe(3)O(4)-ZIF-8 nano-composites for hepatocellular carcinoma therapy. *J Biomater Appl*. May 2019;33(10):1373-1381. doi:10.1177/0885328219836540
48. Tanaka T, Narazaki M, Kishimoto T. IL-6 in inflammation, immunity, and disease. *Cold Spring Harb Perspect Biol*. Sep 4 2014;6(10):a016295. doi:10.1101/cshperspect.a016295
49. Bradley JR. TNF-mediated inflammatory disease. *J Pathol*. Jan 2008;214(2):149-60. doi:10.1002/path.2287

50. Dinarello CA. A clinical perspective of IL-1β as the gatekeeper of inflammation. *Eur J Immunol*. May 2011;41(5):1203-17. doi:10.1002/eji.201141550
51. Soares MP, Marguti I, Cunha A, Larsen R. Immunoregulatory effects of HO-1: how does it work? *Curr Opin Pharmacol*. Aug 2009;9(4):482-9. doi:10.1016/j.coph.2009.05.008
52. Drummond GS, Baum J, Greenberg M, Lewis D, Abraham NG. HO-1 overexpression and underexpression: Clinical implications. *Arch Biochem Biophys*. Sep 30 2019;673:108073. doi:10.1016/j.abb.2019.108073
53. Kaur A, Kumar V, Singh S, et al. Toll-like receptor-associated keratitis and strategies for its management. *3 Biotech*. Oct 2015;5(5):611-619. doi:10.1007/s13205-015-0280-y
54. Olarotimi OJ, Gbore FA, Oloruntola OD, Jimoh OA. Serum inflammation and oxidative DNA damage amelioration in cocks-fed supplemental Vernonia amygdalina and zinc in aflatoxin B(1) contaminated diets. *Transl Anim Sci*. 2023;7(1):txad113. doi:10.1093/tas/txad113
55. Li L, Yuan S, Lin L, et al. Discovery of novel 2-aryl-4-bis-amide imidazoles (ABAI) as anti-inflammatory agents for the treatment of inflammatory bowel diseases (IBD). *Bioorg Chem*. Mar 2022;120:105619. doi:10.1016/j.bioorg.2022.105619
56. Anderson TM, Clay MC, Cioffi AG, et al. Amphotericin forms an extramembranous and fungicidal sterol sponge. *Nat Chem Biol*. May 2014;10(5):400-6. doi:10.1038/nchembio.1496
57. Perfect JR. The antifungal pipeline: a reality check. *Nat Rev Drug Discov*. Sep 2017;16(9):603-616. doi:10.1038/nrd.2017.46
58. Lee Y, Puumala E, Robbins N, Cowen LE. Antifungal Drug Resistance: Molecular Mechanisms in Candida albicans and Beyond. *Chem Rev*. Mar 24 2021;121(6):3390-3411. doi:10.1021/acs.chemrev.0c00199
59. Ravinayagam V, Rehman S. Zeolitic imidazolate framework-8 (ZIF-8) doped TiZSM-5 and Mesoporous carbon for antibacterial characterization. *Saudi J Biol Sci*. Jul 2020;27(7):1726-1736. doi:10.1016/j.sjbs.2020.05.016
60. Liu T, Huang K, Yang Y, et al. An NIR light-driven AgBiS(2)@ZIF-8 hybrid photocatalyst for rapid bacteria-killing. *J Mater Chem B*. Apr 3 2024;12(14):3481-3493. doi:10.1039/d3tb02285d
61. Wang C, Feng S, Tang H, et al. UV-Blocking and Light-Responsive Poly (ε-Caprolactone)/ZIF-8 Multifunctional Composite Films for Efficient Antibacterial Activities. *Chemistry*. Jun 27 2023;29(36):e202300785. doi:10.1002/chem.202300785
62. Tecchio C, Micheletti A, Cassatella MA. Neutrophil-derived cytokines: facts beyond expression. *Front Immunol*. 2014;5:508. doi:10.3389/fimmu.2014.00508
63. Ratitong B, Pearlman E. Pathogenic Aspergillus and Fusarium as important causes of blinding corneal infections - the role of neutrophils in fungal killing, tissue damage and cytokine production. *Curr Opin Microbiol*. Oct 2021;63:195-203. doi:10.1016/j.mib.2021.07.018
64. Ahsan SM, Rao CM. Condition responsive nanoparticles for managing infection and inflammation in keratitis. *Nanoscale*. Jul 20 2017;9(28):9946-9959. doi:10.1039/c7nr00922d
65. Javanbakht S, Hemmati A, Namazi H, Heydari A. Carboxymethylcellulose-coated 5-fluorouracil@MOF-5 nano-hybrid as a bio-nanocomposite carrier for the anticancer oral delivery. *Int J Biol Macromol*. Jul 15 2020;155:876-882. doi:10.1016/j.ijbiomac.2019.12.007
66. Parsaei M, Akhbari K. MOF-801 as a Nanoporous Water-Based Carrier System for In Situ Encapsulation and Sustained Release of 5-FU for Effective Cancer Therapy. *Inorg Chem*. Apr 18 2022;61(15):5912-5925. doi:10.1021/acs.inorgchem.2c00380
67. Ding B, Chen H, Tan J, et al. ZIF-8 Nanoparticles Evoke Pyroptosis for High-Efficiency Cancer Immunotherapy. *Angew Chem Int Ed Engl*. Mar 1 2023;62(10):e202215307. doi:10.1002/anie.202215307
68. Liu W, Tian X, Gu L, et al. Oxymatrine mitigates Aspergillus fumigatus keratitis by suppressing fungal activity and restricting pyroptosis. *Exp Eye Res*. Mar 2024;240:109830. doi:10.1016/j.exer.2024.109830

**Scheme 1** Schematic diagram of the synthesis of ZIF-8 and its effects in vitro and in vivo.

**Figure 1** Characteristic of ZIF-8. **(A)** SEM image with 200nm scale of ZIF-8. **(B)** SEM image with 1µm scale of ZIF-8. **(C-F)** EDS elemental mapping of ZIF-8. **(G)** Distribution histogram of 92 particles. **(H)** Zeta Potential of ZIF-8. **(I)** Powder X-ray diffraction (PXRD) picture of ZIF-8.

**Figure 2** The biological toxicity of ZIF-8 in vitro and in vivo. **(A-B)** Cell viability of RAW 264.7 cells and HCECs in the presence of different concentrations of ZIF-8. **(C-D)** Photographs of Corneal fluorescein staining (CFS) of mice corneas and scores. (ns, significance, **** $p<0.0001$)

**Figure 3** Anti-inflammatory effect of ZIF-8 in RAW 264.7 cells. ZIF-8 decreases the expression of pro-inflammatory factors **(A)** IL-6, **(B)** IL-1β, **(C)** TNF-α, **(D)** TLR-4, and Inflammasome **(E)** NLRP3 mRNA and increases the expression of anti-inflammatory factor **(F)** HO-1 in RAW 264.7 cells. (*** $p<0.001$, **** $p<0.0001$)

**Figure 4** Anti-fungal activity of ZIF-8 against *Aspergillus fumigatus*. **(A)** MIC of ZIF-8 for *A. fumigatus*. **(B)** Inhibitory effect of ZIF-8 on biofilm. Hyphae treated with **(C)** PBS and **(D)** ZIF-8. (ns, no significance, **** $p<0.0001$)

**Figure 5 (A)** PI staining reflects the disruptive effects of ZIF-8 and natamycin on hyphae cell membranes. **(B)** Calcofluor White Staining reveals the impact of ZIF-8 and natamycin on the hyphal cell wall.

**Figure 6** ZIF-8 therapeutic approach ameliorates the prognosis of FK in mice. **(A)** Photographs of corneas of FK-infected mice after the ZIF-8 regimen and **(B)** scoring and corneas of *Aspergillus fumigatus* infected mice after natamycin management. **(C)** Plates of viable fungus in the cornea 3 days after infection. **(D)** Quantitative analysis of fungal load. (ns ,no significance; * $p<0.05$, *** $p<0.001$, **** $p<0.0001$)

**Figure 7** Pathological changes of corneas after ZIF-8 medical intervention. **(A)** HE staining of corneas of different groups of *Aspergillus fumigatus* keratitis mice 3 days after infection (magnification: 400×). **(B)** Immunofluorescence staining of PBS and ZIF-8-treated neutrophils (magnification: 400×).

**Figure 8** The effect of ZIF-8 on reducing inflammatory proteins in RAW 264.7 cells and mice corneas. ZIF-8 decreases the expression of pro-inflammatory proteins **(A)** TNF-α, **(B)** IL-1β, and **(C)** IL-6 in cells after hyphae stimulation. ZIF-8 reduces the expression of proinflammatory proteins **(D)** TNF-α, **(E)** IL-1β, and **(F)** IL-6 in the cornea of mice infected with *Aspergillus fumigatus* 3 days after infection (ns, no significance, **** $p<0.0001$).